\documentstyle [preprint,prl,aps] {revtex}

\begin {document}
\draft
\tightenlines

\title{Application of a new Tight-Binding method for transition
metals:  Manganese}

\author{Michael J. Mehl and Dimitrios A. Papaconstantopoulos}
\address{Complex Systems Theory Branch,Naval Research
Laboratory,Washington, D.C. 20375-5345 USA}

\date{January 27, 1995}
\maketitle

%PACS in Europhysics Letters style format
{}~\\[0.1in]


techniques.

solids.

\begin{abstract}
A new tight-binding total energy method, which has been shown to
accurately predict ground state properties of transition and noble
metals, is applied to Manganese, the element with the most complex
ground state structure among the $d$ metals.  We show that the
tight-binding method correctly predicts the ground state structure
of Mn, and offers some insight into the magnetic properties of this
state.
\end{abstract}

{~}\\[0.1in]
(Submitted to {\em Europhysics Letters})

{~}\\[0.1in]
Most elements in the periodic table crystallize in the fcc, bcc, hcp
and diamond structures.  Among the few exceptions is Manganese,
which has an equilibrium structure, denoted $\alpha$-Mn, which
contains 29 atoms in the unit cell \cite{donohue74}.  First
principles total energy methods, such as the full-potential
Linearized Augmented Plane Wave (LAPW) method
\cite{andersen75,wei85} are not very efficient in such systems,
especially since the $\alpha$-Mn phase has five internal parameters
which must be adjusted to minimize the total energy in order to
calculate the correct structure at each volume.  Calculation of
elastic constants, phonon frequencies, surface energies, vacancy
formation energies, and other properties require even more
computational effort.  A reliable approximate method, based on
first-principles results, is necessary for efficient computational
study of complicated crystals such as Manganese.

In a recent paper, Sigalas and Papaconstantopoulos \cite{sigalas94a}
introduced the idea that the energy bands of Augmented Plane Wave
(APW) calculations for cubic structures at different volumes can be
fit to a non-orthogonal tight-binding (TB) Hamiltonian whose matrix
elements are functions of the distance between pairs of atoms.  The
sum of eigenvalues resulting from the above TB Hamiltonian, together
with a pair potential, were used to fit the total energies of the
APW calculation, thus obtaining an interpolation formula that was
employed to calculate the total energy for non-cubic structures.
This procedure was applied to calculate the elastic constants of Pd,
Ir, Au, Rh and Ta, which showed fairly good agreement with the
experimental values.  The phonon spectra and density of states for
Au were also calculated \cite{mehl94}, again in reasonable agreement
with experiment.

In a subsequent paper, Cohen, Mehl and Papaconstantopoulos
\cite{cohen94} made dramatic improvements to the above approach.
They eliminated the pair potential in the fitting of the total
energy; employed environment-dependent on-site TB parameters; and
introduced exponentially damped polynomial expansions of the hopping
and overlap integrals, thus extending the parametrization to an
arbitrary number of neighbors.  This new total energy methodology
\cite{cohen94} was applied to calculate elastic constants, phonon
spectra and vacancy formation energies for the noble metals and
other transition metals.  The results were impressive.  Starting
{}from only fcc and bcc structures, the method correctly predicted the
ground state structure in all of the elements tested, including
those which exhibit an hexagonal close-packed (hcp) ground state.

This paper shows how the new TB method can be applied to Manganese.
We first performed paramagnetic LAPW calculations at five volumes in
each of the monatomic fcc and bcc structures, and then determined a
set of TB parameters \cite{cohen94} which reproduced the electronic
structure and total energy of these structures.  We then used the
resulting Hamiltonian to compute the total energy of Manganese in
the $\alpha$-Mn, $\beta$-Mn \cite{donohue74,shoemaker78}, fcc, bcc,
hcp, and simple cubic (sc) structures.  The $\alpha$-Mn structure
has a bcc unit cell containing atoms, with five internal parameters.
Its space group is $I\overline{4}3m\mbox{-}T^3_d$\cite{donohue74}.
The twenty-nine atoms are divided up into four types, with all atoms
of a given type equivalent by symmetry.  There is one atom of type
$I$, located at the origin; four atoms of type $II$, located at
$a(x_1,x_1,x_1)$ and equivalent points, where $a$ is the cubic
lattice constant; twelve atoms of type $III$, located at
$a(x_2,x_2,z_2)$ and equivalent points, and twelve atoms of type
$IV$, located at $a(x_3,x_3,z_3)$ and equivalent points.  The
$\beta$-Mn structure has a simple cubic structure containing twenty
atoms, with two internal parameters.  The space group is
$P4_332\mbox{-}O^6$\cite{donohue74}.  There are eight atoms of type
$I$, located at $a (x_1,x_1,x_1)$ and equivalent sites; and twelve
atoms of type $II$, located at $a(1/8, x_2, 1/4 + x_2)$ and
equivalent sites.  Neither structure has an inversion site, so we
must solve the generalized eigenvalue problem for Hermitian
matrices.  The large number of atoms, coupled with the necessity of
minimizing the total energy with respect to the internal parameters,
makes the determination of structural properties difficult to handle
by first-principles total energy electronic structure calculations.
Within the tight-binding method, however, the calculation is
relatively easy.  In Table~\ref{tab:eos} we show the equilibrium
volume, relative energy, and bulk modulus as calculated by our TB
procedure for Manganese, as well as Technetium and Copper for
comparison.  Fig.~\ref{fig:mneos} shows the energy/volume
relationship for several of the lower energy phases, and
Table~\ref{tab:mnpar} compares the equilibrium structural parameters
with the experimental ones.  Our TB Hamiltonian correctly predicts
the ground state structure of Manganese.  The calculation shows that
the $\beta$-Mn phase is close in energy to the $\alpha$-Mn phase,
indicating that it is a likely candidate for the high-temperature
phase of Manganese, in agreement with experiment \cite{donohue74}.
Our calculations also predict that $\alpha$-Mn will transform into
an hcp structure at a pressure of 50 GPa.

Of course this agreement with experiment could be an artifact of the
way we constructed the Hamiltonian.  To test this, we constructed TB
Hamiltonians for Technetium, which is also a column VIIB element,
and Copper by determining Tight-Binding parameters which reproduced
the results of APW (for Technetium) and LAPW (for Copper)
calculations as outlined above.  The resulting equation of state
data is presented in Table~\ref{tab:eos}.  Our parametrization
correctly predicts that the hcp phase is the ground state for
Technetium although the $\alpha$-Mn and $\beta$-Mn phase are close
in energy.  Copper is correctly predicted to be in the fcc phase,
while the $\alpha$-Mn and $\beta$-Mn phases are respectively 6.0 mRy
and 5.4 mRy higher than the fcc phase.

Our calculations for Manganese were performed assuming a
paramagnetic phase, while experiment \cite{landolt} and theory
\cite{sliwko94,fuster88,oguchi84} suggest that most phases of
Manganese exhibit some form of magnetism, and that the $\alpha$
phase is antiferromagnetic \cite{landolt,sliwko94}.  This does not
affect the fact that our method demonstrates that the $\alpha$ phase
is the ground state of Manganese, because the addition of magnetism
can only lower the energy of the $\alpha$ phase.  Magnetism will,
however, affect the volume of the ground state phase.  Our
equilibrium lattice constant is about $6\%$ smaller than the
experimental lattice constant.  This error can be partially
attributed to the neglect of magnetism and partially to the error
inherent in the LDA \cite{moruzzi78}.  Our calculated internal
parameters (Table~\ref{tab:mnpar}) are almost identical to the
experimentally measured parameters, so we conclude that the internal
parameters are not changed by magnetism.

Since the current formalism is not set up for spin-polarized
calculations, we use our paramagnetic TB Hamiltonian to calculate
the electronic density of states (DOS) for the $\alpha$-Mn phase.
Fig.~\ref{fig:mndos} shows the total DOS of $\alpha$-Mn as well as
the $d$ partial DOS for the four different atom types.  The width of
the $d$ states appears the same for all sites.  However, there are
differences in the details of the DOS structure.  In particular, the
Fermi level values of the DOS differ substantially.  This is shown
in Table~\ref{tab:partdos} where we note that the first two sites
have DOS values which are a factor of two larger than the other two
sites.  Also from Table~\ref{tab:partdos} we note that the $s$ and
$p$-like DOS at $E_F$ are very small.  We then applied the Stoner
criterion \cite{vosko75,janak77} using the DOS at the Fermi level in
conjunction with a matrix element derived from fcc and bcc
calculations \cite{sigalas94b}.  Using an approximate value of $I_F
= 0.03 Ry$ \cite{sigalas94b}, we obtained values of the Stoner $nI$
of about 0.7 for atoms on sites $I$ and $II$, and about 0.4 for
atoms on sites $III$ and $IV$.  This is consistent with
first-principles band-structure calculations for the moments on the
atoms \cite{sliwko94}, where it is found that atoms $I$ and $II$
have large moments, but atoms $III$ and $IV$ have smaller moments.

We conclude that our TB total energy method is capable of predicting
the correct total energy ordering of various structures, including
the complicated $\alpha$-Mn structure, with computational costs
orders of magnitude lower than standard first-principles
calculations.  In addition this scheme provides reliable energy
bands and DOS for all phases of Manganese.

We wish to thank R. Cohen, W. Pickett, and D. Singh for useful
discussions.

\begin{figure}
\caption{The total energy of Manganese in several structures as a
function of volume, obtained from the tight-binding Hamiltonian
outlined in the text.  We show energy per atom versus volume per
atom for ease in comparison.  The $\diamond$ symbols indicate the
fcc and bcc phase LAPW energies used in the fit.  The $\alpha$-Mn
phase is correctly predicted to be the ground state.}
\label{fig:mneos}
\end{figure}

\begin{figure}
\caption{The electronic density of states (DOS) of the $\alpha$
phase of Manganese, as well as the $d$ density of states for the
four distinct sites.  The dotted vertical line represents the Fermi
energy.}
\label{fig:mndos}
\end{figure}

\begin{table}
\caption{The equilibrium volume ($V_0$) in Bohr$^3$ and energy
($E_0$) in $mRy$ per atom, and bulk modulus ($B_0$) in GPa for
Manganese, Technetium, and Copper, as calculated by the
Tight-Binding method.  $E_0$ is set to zero for the ground state
energy of each element.}
\begin{tabular}{ccccccccc}
Atom &  & fcc & bcc & hcp & sc & diamond & $\alpha$-Mn & $\beta$-Mn
\\
\tableline
Mn & $V_0$ &  68.7 &  68.7 &  68.4 &  74.0 &
 96.8 &  69.3 &  69.6 \\
   & $E_0$ &  8.2  &  15.5 &   3.1 &  91.0 &
172.0 &   0.0 &   1.6 \\
   & $B_0$ &  315  &  324  &  314  &  199  &
 119  &  320  &  318  \\
\tableline
Tc & $V_0$ &  94.2 &  95.4 &  93.6 &  102.5 &
127.2 &  95.1 &  95.1 \\
   & $E_0$ &   6.5 &  23.9 &   0.0 &   57.2 &
 73.2 &   0.2 &   2.6 \\
   & $B_0$ &  309      &  306      &  303      &  244       &  175
      &  299      &  298      \\
\tableline
Cu & $V_0$ &  73.6 &  73.9 &  73.8 &  82.6 &
106.5 &  75.2 &  75.5 \\
   & $E_0$ &   0.0 &   3.5 &   1.5 &  25.1 &
 70.9 &   6.0 &   5.4 \\
   & $B_0$ &    190    &    186    &    186    &    141    &     64
      &    176    &    177    \\
\end{tabular}
\label{tab:eos}
\end{table}

\begin{table}
\caption{The experimental and TB equilibrium lattice and internal
parameters for the $\alpha$ structure of Manganese.  The bcc unit
cell has twenty-nine atoms, divided into four classes as explained
in the text.}
\begin{tabular}{lcccccc}
	& $a$(\AA) & $x_1$ & $x_2$ & $z_2$ & $x_3$ & $z_3$ \\
\tableline
Experiment &  8.9129 &  0.31765 &  0.35711 &
0.03470 &  0.08968 &  0.28211 \\
Tight-Binding &  8.41 &  0.31719 &  0.35787 &
0.03964 &  0.08971 &  0.27983 \\
\end{tabular}
\label{tab:mnpar}
\end{table}

\begin{table}
\caption{The total electronic density of states at the Fermi level
for $\alpha$-Mn at the minimum energy volume predicted by the
tight-binding calculations.  The partial DOS for the $s$, $p$, and
$d$ states are shown for each atom type (see text).  The Stoner
criterion parameter is calculated assuming $I_F = 0.03$ Ry
\protect\cite{sigalas94b}.  The coordination numbers are those
assigned by Donohue\protect\cite{donohue74}.}
\begin{tabular}{cccccccc}
\multicolumn{3}{c}{Atom} & \multicolumn{4}{c}{DOS at $E_F$} &
Stoner \\
 Type & Number & Coordination & Total & \multicolumn{3}{c}{Partial}
& Criterion \\
 & & & (States/Ry/Unit Cell) & \multicolumn{3}{c}{(States/Ry/Atom)}
& \\
 & & & & $s$ & $p$ & $d$ & \\
\tableline
 I & 1 & 12 & 458.39706 & 0.07778 & 0.70738 & 22.81580 & 0.68 \\
II & 4 & 10 &  & 0.27634 & 0.43595 & 25.07444 & 0.75 \\
III & 12 & 13 & & 0.21981 & 0.64905 & 14.41710 & 0.43 \\
IV & 12 & 11 & & 0.13672 & 0.60299 & 11.61058 & 0.35 \\
\end{tabular}
\label{tab:partdos}
\end{table}

\end {document}
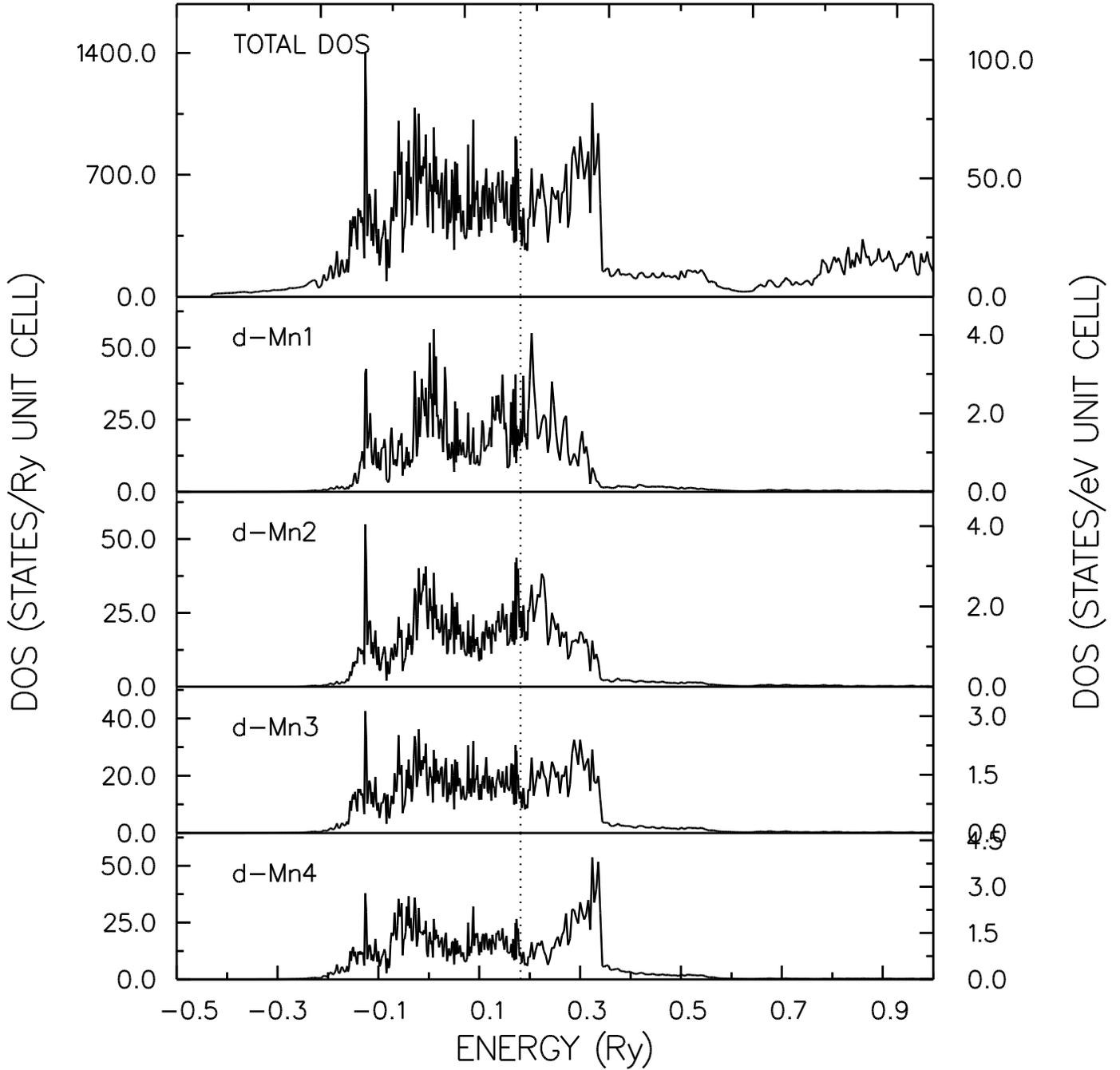
